\documentclass[twocolumn]{aastex631}

\usepackage{multirow}
\usepackage{float}

\begin{document}
\title{First measurement of $^{87}$Rb($\alpha,xn$) cross sections at weak $r$-process energies in supernova $\nu$-driven ejecta to investigate elemental abundances in low-metallicity stars}

\author[0000-0002-1236-4739]{C. Fougères}
\affiliation{Physics Division, Argonne National Laboratory 
Lemont, IL 60439, USA}
\affiliation{CEA, DAM, DIF, 91297 Arpajon, France }
\affiliation{Laboratoire Matière en Conditions Extr\^{e}mes, Université Paris-Saclay, CEA, 91680~Bruyères-le-Ch\^{a}tel,~France}
\author[0009-0002-4051-9627]{M. L. Avila}
\affiliation{Physics Division, Argonne National Laboratory 
Lemont, IL 60439, USA}
\author[0000-0003-2197-0797]{A. Psaltis}
\affiliation{Department of Physics, Duke University, Durham, NC, 27710, USA}
\affiliation{Triangle Universities Nuclear Laboratory, Duke University, Durham, NC, 27710, USA}
\author[0000-0002-3763-714X]{M. Anastasiou}
\affiliation{Nuclear and Chemical Sciences Division, Lawrence Livermore National Laboratory, Livermore, CA 94550, USA}
\author{S. Bae}
\affiliation{Center for Exotic Nuclear Studies, Institute for Basic Science, Daejeon, 34126, Korea}
\author{L. Balliet}
\affiliation{Facility for Rare Isotope Beams (FRIB), Michigan State University, East Lansing, MI 48824, USA}
\author{K. Bhatt}
\affiliation{Physics Division, Argonne National Laboratory 
Lemont, IL 60439, USA}
\author{L. Dienis}
\affiliation{Grand Accélérateur National d’Ions Lourds (GANIL), Caen, France}
\author[0000-0001-8746-0234]{ H. Jayatissa}
\affiliation{Physics Division, Los Alamos National Laboratory, NM 87545, USA}
\author[0000-0002-6497-0175]{V. Karayonchev}
\affiliation{Physics Division, Argonne National Laboratory 
Lemont, IL 60439, USA}
 \author[0000-0002-6695-9359]{P. Mohr}
    \affiliation{HUN-REN Institute for Nuclear Research (ATOMKI), H-4001 Debrecen, Hungary}
\author[0000-0001-9849-5555]{ F. Montes}
\affiliation{Facility for Rare Isotope Beams (FRIB), Michigan State University, East Lansing, MI 48824, USA}
\author[0000-0002-5397-7048]{D. Neto}
\affiliation{Physics Division, Argonne National Laboratory 
Lemont, IL 60439, USA}
\affiliation{Department of Physics, University of Illinois Chicago, 845 W. Taylor St., Chicago, IL 60607, USA}
\author[0000-0003-4539-5985]{F. de Oliveira Santos}
\affiliation{Grand Accélérateur National d’Ions Lourds (GANIL), Caen, France}
\author{W.-J. Ong}
\affiliation{Nuclear and Chemical Sciences Division, Lawrence Livermore National Laboratory, Livermore, CA 94550, USA}
\author[0009-0009-4827-0863]{ K. E. Rehm}
\affiliation{Physics Division, Argonne National Laboratory 
Lemont, IL 60439, USA}
\author{W. Reviol}
\affiliation{Physics Division, Argonne National Laboratory 
Lemont, IL 60439, USA}
\author[0000-0003-3125-9907]{D. Santiago-Gonzalez}
\affiliation{Physics Division, Argonne National Laboratory 
Lemont, IL 60439, USA}
\author[0000-0002-5046-9451]{N. Sensharma}
\affiliation{Physics Division, Argonne National Laboratory 
Lemont, IL 60439, USA}
\author[0000-0002-1637-7502]{R. S. Sidhu}
\affiliation{School of Mathematics and Physics, University of Surrey, Guildford, GU2 7XH, United Kingdom}
\author[0000-0002-6631-7479]{I. A. Tolstukhin}
\affiliation{Physics Division, Argonne National Laboratory 
Lemont, IL 60439, USA}

\begin{abstract}
Observed abundances of $Z\sim40$ elements in metal-poor stars vary from star to star, indicating that the rapid and slow neutron capture processes may not contribute alone to the synthesis of elements beyond iron. The weak $r$-process was proposed to produce $Z\sim40$ elements in a subset of old stars. Thought to occur in the $\nu$-driven ejecta of a core-collapse supernova, ($\alpha,xn$) reactions would drive the nuclear flow toward heavier masses at $T=2-5$~GK. However, current comparisons between modelled and observed yields do not bring satisfactory insights into the stellar environment, mainly due to the uncertainties of the nuclear physics inputs where the dispersion in a given reaction rate often exceeds one order of magnitude. Involved rates are calculated with the statistical model where the choice of an $\alpha$-optical-model potential ($\alpha$OMP) leads to such a poor precision. The first experiment on $^{87}$Rb($\alpha,xn$) reactions at weak $r$-process energies is reported here. Total inclusive cross sections were assessed at $E_{c.m.}=8.1-13$~MeV ($3.7-7.6$~GK) with the active target MUlti-Sampling Ionization Chamber (MUSIC). With a $N=50$ seed nucleus, the measured values agree with statistical model estimates using the $\alpha$OMP {\texttt{Atomki-V2}}. A re-evaluated reaction rate was incorporated into new nucleosynthesis calculations, focusing on $\nu$-driven ejecta conditions known to be sensitive to this specific rate. These conditions were found to fail to reproduce the lighter-heavy element abundances in metal-poor stars.
\end{abstract}
\correspondingauthor{C. Fougères}
\email{chloe.fougeres@cea.fr}
\section{Introduction} \label{sec:intro}
\par The oldest, metal-poor, stars in the Milky Way and in close dwarf galaxies have been investigated over the past decade to bring forth the presence of elements heavier than iron. At these sites, the observed chemical abundances~\citep{artFrebel, Cote2019, Reichert2020} hint that the rapid neutron capture process~\citep{NatureRevRprocc}, expected to produce half of nuclei beyond Fe, should take place in early galactic ages. An active site of this nucleosynthesis has been recently found with the observation of $r$-process elements in the kilonova following a binary neutron star merger (NSM)~\citep{Kasen17, Smartt17, NatSrkilonova}. However, it occurs in galactic evolution too rarely and too late to explain the observed chemical abundances in old stars~\citep{Cote2019, Kobayashi2023}. Other explosive stellar environments like magnetorotationally-driven supernovae~\citep{Winteler2012, Nishimura, Reichert2021, Reichert2023a}, collapsars~\citep{2019Natur.569..241S}, neutrino($\nu$)-driven winds in core-collapse supernovae (CCSNe)~\citep{Hansen14,Horowitz19} are being investigated.
\par  An enhancement of the elements around the first $r$-process peak, with $Z=38–47$,  was also observed in a subset of metal-poor stars~\citep{lighAbund, refMashonkina, Schwerdtfeger}. These abundance patterns, e.g.~Fig.~7~\citep{Schwerdtfeger} and Tables 2~\citep{Psaltis2022, Psaltis2024}, call for an additional mechanism that must occur in early galactic ages. Two have been put forward:~the weak $r$-process ~\citep{Montes2007, QIAN2007237, 2009ApJ...692.1517I,  Arcones11, Bliss,Bliss2018,Bliss2020, Psaltis2022} also known as the $\alpha$-process~\citep{1992ApJ...399..656M}, and the neutrino-proton process ($\nu p$-process)~\citep{PhysRevLett.96.142502,  Nishimura2019}. Both are expected to take place in $\nu$-driven ejecta of stellar explosions like CCSNe or NSMs, depending on the neutron-richness of the ejected material. We focus here on the weak $r$-process in $\nu$-driven winds of core-collapse supernovae.
\par In the aftermath of the collapse, temperature decreases while the neutron-rich matter is expanding away from the compact neutron star and, at some point, the nuclear statistical equilibrium breaks down. Regardless of the thermodynamics conditions (expansion time scale, entropy, electron fraction) in these extreme winds, parametric 1-dimension modelling studies~\citep{Bliss,Bliss2018,Bliss2020} show that $(i)$ radiative $n$-capture reactions are balanced by their reciprocal photo-disintegration reactions, $(ii)$ $\beta$ decays occur more slowly than the expansion time scale of tens of ms, and $(iii)$ $(\alpha,1n)$ and  $(\alpha,2n)$ reactions are faster to fall out of equilibrium than other $\alpha$-induced and $p$-induced reactions at temperatures of $2 - 5$~GK. Hence, the nucleosynthesis pathway should stay relatively close to stability, and the nuclear reaction flow toward the elements from Fe to Mo should be driven by $(\alpha,xn)$ reactions where $x=1,2$ are the typical cases. Several sensitivity studies~\citep{Bliss2018, Bliss2020, Psaltis2022, Psaltis2024} have shown that model-to-observations comparisons of abundances around Sr in metal-poor stars are currently inadequate to firmly constrain the thermodynamics conditions of the $\nu$-driven ejecta. This is mainly due to the variations resulting from the uncertainties of ($\alpha,xn$) reaction rates which have been poorly measured so far.

\par Without experimental information, ($\alpha,xn$) reaction rates are estimated within the Hauser-Feshbach (HF) framework~\citep{PhysRev.87.366, RAUSCHER20001}. This model is justified for the mass region of interest and the involved stellar temperatures ($T>1$~GK) which correspond to the energy region of high nuclear level density in the compound nucleus. Since the decay of the latter is independent of its formation mechanism, the probability that the ($\alpha,xn$) reaction occurs can be expressed as $\sigma(\alpha,xn) \approx  T_{\alpha,0} \frac{T_{xn}}{\sum T_i}$, i.e.~the product of the transmission coefficient of the $\alpha$ particle into the seed nucleus ($T_{\alpha,0}$) and the transmission coefficient of the $xn$-exit channel ($T_{xn}$) that is normalized to all exit channels (${\sum T_i}$). The open exit channels  are $\gamma$ rays and multiple particles ($p$, $n$,  $2n$, $\alpha$,...). However, at weak $r$-process energies located far above the $n$ emission threshold, $xn$-exit channels dominate and, so, the $\frac{T_{xn}}{\sum T_i}$ term reduces to 1. Hence, only the $T_{\alpha,0}$ coefficient is relevant for statistical model estimates of $(\alpha,xn)$ cross sections.
\par
\begin{figure}[ht!]
\begin{center}
\includegraphics[scale=0.44]{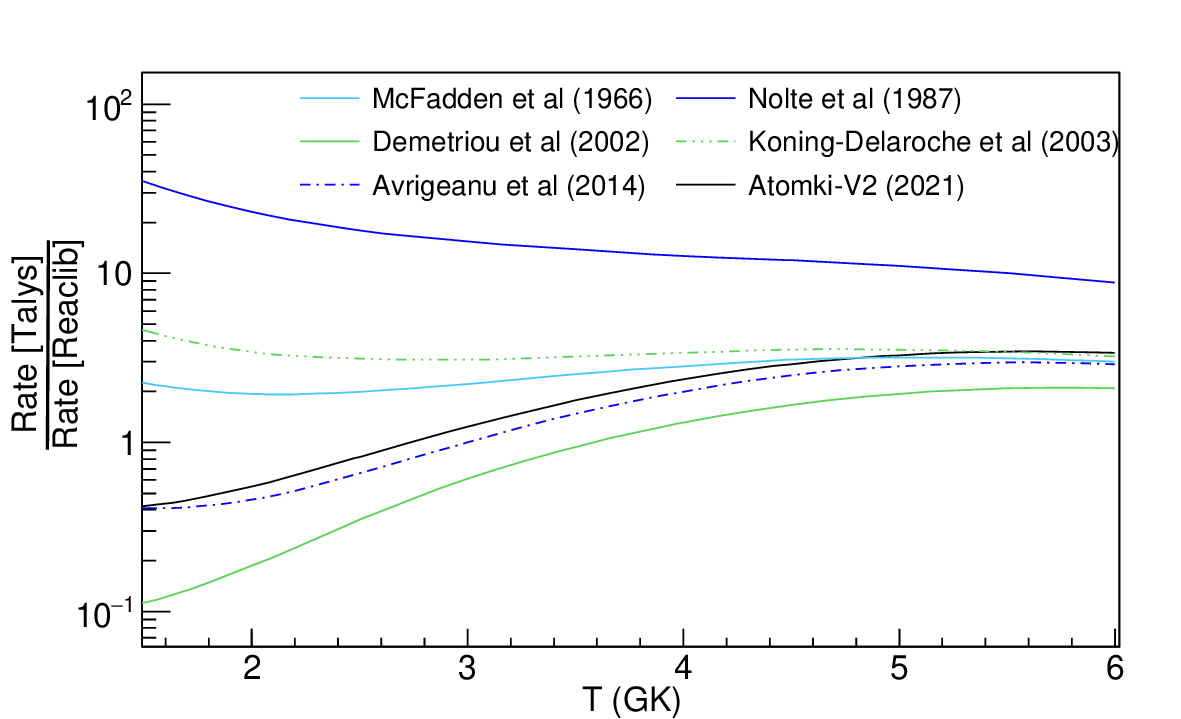}
\end{center}
\caption{Ratio between the $^{87}$Rb($\alpha,1n$)$^{90}$Y reaction rate calculated with the {\texttt{Talys}} code~\cite{talys2, talys1}  and the rate from {\texttt{ReaclibV2.2}}~\cite{Cyburt2010} along temperatures where ($\alpha,xn$) reactions impact the nucleosynthesis occuring in CCSNe $\nu$-driven winds. Standard $\alpha$OMPs were considered to calculate the reaction rate, i.e.~those from~\cite{aOMPmcfadden} shown in solid cyan,  from~\cite{aOMPnolte} shown in solid blue,  from~\cite{aOMPdemetriou}  with the dispersive mode shown in solid green, from nucleon potential~\cite{aOMPkoning} with the folding approach~\cite{WATANABE1958484} shown in dotted green, from~\cite{Avrigeanu2014} shown in dotted blue, and from \cite{atomkiv2} shown in solid black. The maximal deviations range from a factor 10 at $T=5$~GK to a factor 100 at $T=2$~GK.}
\label{fig:RateFluc}
\end{figure}

 The $T_{\alpha,0}$ coefficient is derived from an $\alpha$-Optical-Model Potential ($\alpha$OMP), see e.g.~\cite{atomkiv2}. Several $\alpha$OMPs are available, to quote the standard ones~\citep{aOMPmcfadden,aOMPnolte,aOMPdemetriou}, \cite{aOMPkoning} with the folding approach~\cite{WATANABE1958484}, \citep{Avrigeanu2014,atomkiv2}. Deviations between reaction rates of one to two orders of magnitude have been observed  while testing different $\alpha$OMPs. This is illustrated in Figure~\ref{fig:RateFluc}  where several $^{87}$Rb($\alpha,1n$)$^{90}$Y reaction rates, normalized to the referenced values of {\texttt{ReaclibV2.2}} \cite{Cyburt2010}, are presented along temperatures relevant for the weak $r$-process. The reaction rates were calculated with the code {\texttt{Talys}}~\citep{talys2, talys1} where only the chosen $\alpha$OMP was varied. Note that the ratio resulting from the $\alpha$OMP~\citep{aOMPmcfadden} (solid cyan curve Figure~\ref{fig:RateFluc}) deviates from unity even though the same $\alpha$OMP was used for the reaction rate calculations in {\texttt{ReaclibV2.2}}~\citep{Cyburt2010}, but with the statistical model code {\texttt{NON-SMOKER}}~\citep{NONSMOKERref} and technical differences exist between the two codes.
\par Measured data allow us to test the statistical model predictions based on available $\alpha$OMPs for $(\alpha,xn)$ cross sections with the goal to improve the precision of nucleosynthesis calculations for the weak $r$-process. In this respect, statistical model estimates based on the latest $\alpha$OMP {\texttt{Atomki-V2}}~\citep{atomkiv2} appeared to be consistent with some recent measurements performed on stable nuclei located in the weak $r$-process mass region. These are~\citep{PhysRevC.78.025804, ref86Sr,PhysRevC.104.035804, Kiss2021, Ong22} for which the deviations between experimental and calculated values of ($\alpha,1n$) cross sections are within a factor $0.5 - 2$. However, a recent study on $^{88}$Sr at $N=50$~\citep{PhysRevC.109.065805} measured  a cross section that is systematically lower ($\sim32$~\%) than statistical model estimates with the $\alpha$OMP {\texttt{Atomki-V2}}. This, together with the general lack of data available on the neutron-rich side to test statistical model predictions, push for more nuclear physics measurements.
\par The comprehensive sensitivity studies of~\cite{Bliss2020, Psaltis2022} have identified the key ($\alpha,xn$) reactions that strongly contribute to the uncertainties of the yields derived from the modelling of CCSNe $\nu$-driven winds. These works have determined  which elemental abundances (and by how much) are affected by a given ($\alpha,xn$) reaction rate under certain thermodynamics conditions in the $\nu$-driven ejecta, and thus suggest a selection of experiments that would best help constrain model-to-observations comparisons for abundances in old stars. Such experimental efforts on measuring ($\alpha,xn$) weak $r$-process cross sections have already started with stable beams~\cite{PhysRevC.104.035804, Kiss2021, Ong22,PhysRevC.109.065805}. Many more, but with unstable neutron-rich beams or targets, have yet to be assessed. Heavy-ion accelerators like ATLAS (US) or FRIB (US) are now enabling such experimental programs at  weak $r$-process energies ($\sim 1 - 3$~MeV/u) thanks to the high intensities available for beams of the desired neutron-rich isotopes.
\par Among the stable cases of astrophysical relevance that remain to be measured at astrophysical energies, the reaction $^{87}$Rb($\alpha,1n$)$^{90}$Y was found to impact $Z=41, 42, 44, 45$ abundances by factors of $2 - 3$ and (Sr/Zr, Y/Zr, Nb/Zr) elemental ratios in a handful of CCSNe $\nu$-driven winds conditions~\citep{Bliss2020, Psaltis2022}. Located at the closed shell $N = 50$, this should further test statistical model predictions with the up-to-date $\alpha$OMP {\texttt{Atomki-V2}}~\cite{atomkiv2}. To our knowledge, $^{87}$Rb($\alpha,1n$)$^{90}$Y cross sections were only measured above the weak $r$-process energy region ($E_{c.m.}\geq10.52$~MeV, $T\geq5.5$~GK) with uncertainties of $\sim20$~\%~\cite{Riley1964}. The reaction $^{87}$Rb($\alpha,2n$)$^{89}$Y open at $T>5.6$~GK has not been investigated at all. An inclusive measurement of the two reactions would directly constrain the $\alpha$OMP since the summed contribution of the ($\alpha,1n$) and ($\alpha,2n$) channels depends solely on the $\alpha$OMP.
\par Responding to the need of data at weak $r$-process energies and of constraints on the $\alpha$OMP, this work presents the first measurement of the total cross sections of both $^{87}$Rb($\alpha,xn$) reactions at $E_{c.m.}=8.1 - 13$~MeV (T$\sim 3.7 - 7.6$~GK). The inclusive excitation function was measured with the active gaseous technique. The measured cross sections were then compared to statistical model estimates and the $^{87}$Rb($\alpha,1n$)$^{90}$Y reaction rate determined in the temperature range relevant for the weak $r$-process in CCSNe $\nu$-driven winds. The impact of this newly constrained reaction rate was assessed on $Z\sim40$ elemental abundances by nucleosynthesis calculations in CCSNe $\nu$-driven ejecta presenting thermodynamics conditions known to be sensitive to this specific rate. The resulting predictions were finally compared to the abundances observed in a set of old stars.

\section{Nuclear experiment} \label{sec:exp}
\par Earlier experimental investigations on $(\alpha,1n$) reactions have tested the standard $\alpha$OMPs in the mass region relevant for the weak $r$-process~\citep{PhysRevC.104.035804, Ong22}, including at $N=50$ closed-shell nuclei~\citep{refConf86Kr,PhysRevC.109.065805}, and reported constraints on modelled elemental abundances in CCSNe $\nu$-driven winds~\citep{Kiss2021}. The stellar reaction rates are determined for charged-particle induced reactions at energies $\simeq$1~MeV above the neutron emission threshold, and, so, most attempts aim at directly assessing high cross sections ($0.1-100$~mb). To quote recent works, this can be achieved through $(i)$ integrated measurements with the active target technique in inverse kinematics~\citep{Ong22, PhysRevC.109.065805}, or $(ii)$  $\beta-$delayed-$\gamma$-ray measurements with the activation technique in normal kinematics~\citep{ref86Sr, PhysRevC.104.035804,Kiss2021}, or  $(iii)$ recoil-$\gamma$-ray prompt coincidence measurements at a mass separator using direct reactions in inverse kinematics~\citep{refConf86Kr}. The present work on $^{87}$Rb$(\alpha,xn$) reactions was carried out using the experimental technique $(i)$.

\subsection{Method}
\par The active target technique relies on the detection medium being also the target material. Cross sections are straightforwardly measured while detecting the nuclei of both the entrance and exit channels of the reaction.  This technique presents many advantages:~an increased target thickness, a detection efficiency of $\sim100$~\%, a self-normalization of the cross section. In the case of $\alpha$-induced reactions, the technique can be employed in inverse kinematics with active helium-gaseous targets being ionization chambers~\citep{AvilaNIM16,JOHNSTONE2021165697,BLANKSTEIN2023167777}, time-projection chambers~\citep{ACTAR2, KOSHCHIY2020163398,  AYYAD2020161341}, or arrays that combine gaseous and semi-conductor detectors~\citep{KOSHCHIY20171}. The excitation function is probed at different center-of-mass energies ($E_{c.m.}$) while the incident mono-energetic beam is slowing down in the gaseous volume. The reachable range and resolution in $E_{c.m.}$  are governed by incident beam energy, gas pressure, and detector segmentation. Most high-profile seed nuclei in explosive stellar nucleosynthesis are radioactive, short-lived, elements. The inverse kinematics method used here allows us to investigate them thanks to the delivered radioactive beams. 
\par Depending on the detector sensitivity, the experiment focuses on selecting the heavy recoils or light ejectiles as well as following energy losses or reconstructing reaction vertices and kinematics in order to identify the reaction channel of interest. Since energy loss of ions in matter varies as the square of the atomic number, ($\alpha,xn$) reactions on $Z\sim40$ nuclei can be observed via the first method:~the measurement of heavy-ion energy losses as they pass through the active gaseous volume. Due to the $Z+2$ change between the beam and the heavy recoil, a sharp local increase in energy loss is a signature of a reaction occurring. With electrically-segmented (stripped) detectors in a single direction~\citep{MUSICsetup,JOHNSTONE2021165697,BLANKSTEIN2023167777}, the search for ($\alpha,xn$) events may follow:
\begin{itemize}
\item the {\itshape{PID method}}, a global analysis that looks at energy losses summed over several strips $\Delta E - E_{total}$, similar to the Particle IDentification (PID) with silicon detectors and their variants, 
\item the {\itshape{Traces method}}, a local analysis that tracks energy loss per strip.
\end{itemize}
\par It is sufficient to analyze energy losses relative to the averaged value found for the beam in order to identify reaction events. However, the absolute energy losses of the beam particles must be known to determine the center-of-mass energies associated with the reaction of interest. The precision in  the determination of the beam energy losses directly impacts the energy resolution of the measured excitation function. Energy losses of heavy ions in matter can be estimated with Monte Carlo calculations that integrate tabulated stopping powers. However, large deviations ($>$10~\%) have been observed at low energies between tables from standard libraries~\citep{HUBERT19901, srim, ATIMA}. Hence, performing also a measurement of the beam energy losses in the gaseous detector is essential in active-target experiments that aim at assessing cross sections.
\par The technique described here was chosen to determine the $^{87}$Rb($\alpha,xn$) reaction cross sections. Guided by  past successful experiments~\citep{Avila16,AvilaNIM16,  Ong22, PhysRevC.109.065805}, the measurement was carried out in inverse kinematics at ATLAS accelerator with the electrically-segmented MUlti Sampling Ionization Chamber (MUSIC)~\citep{MUSICsetup}.

\subsection{Set-up}
\begin{figure*}[ht!]
\begin{center}
\includegraphics[scale=0.37]{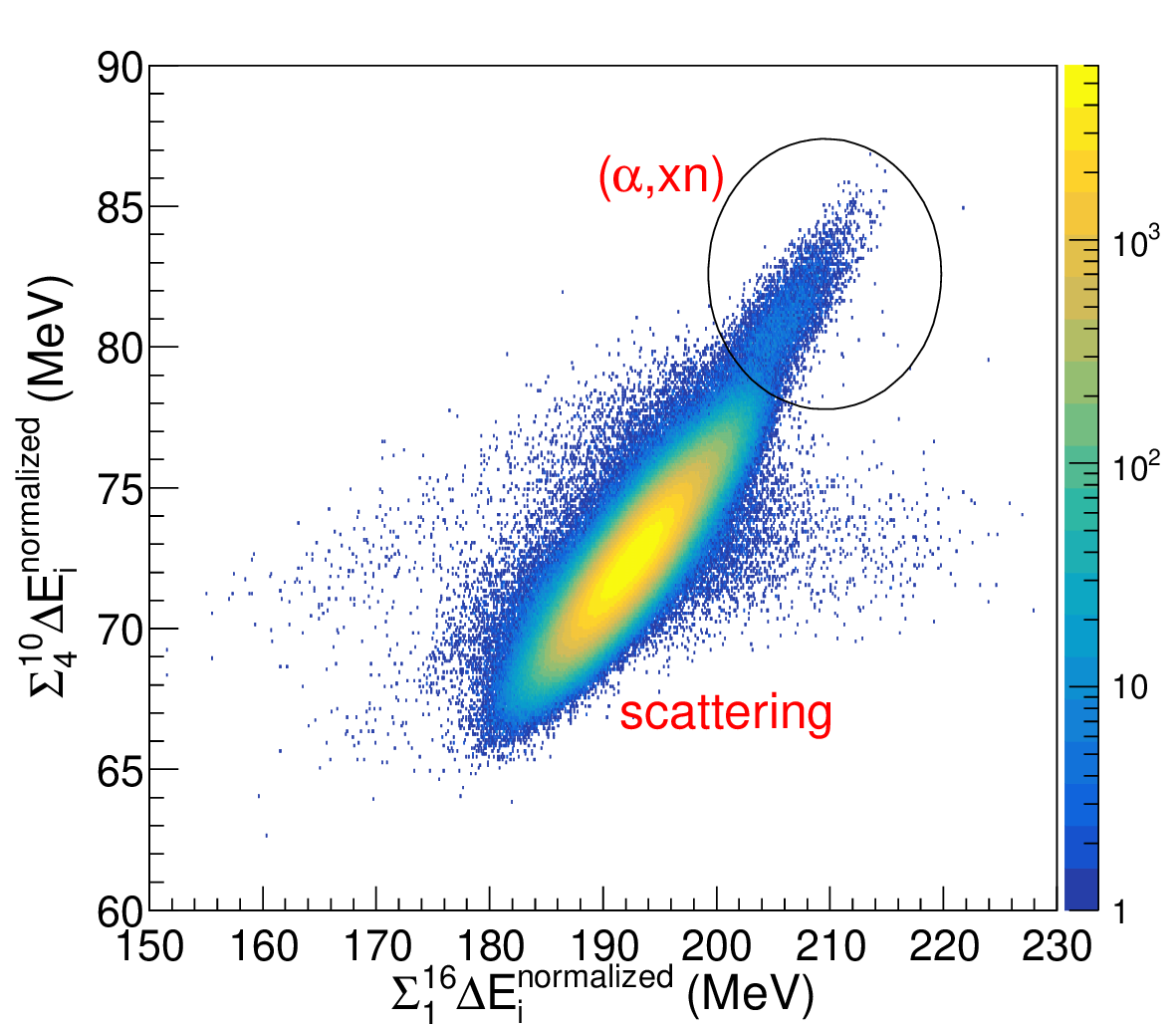}\includegraphics[scale=0.58]{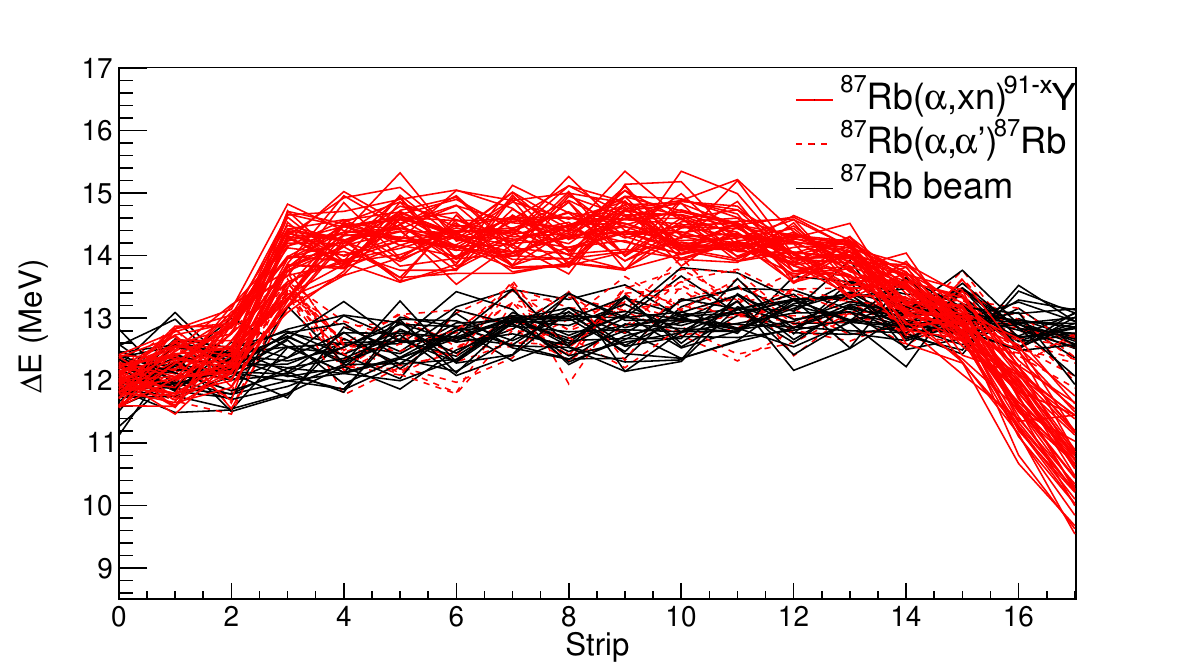}
\end{center}
\caption{Application of the {\itshape{PID method}} (left) and  {\itshape{Traces method}} (right) to identify and quantify ($\alpha,xn$) reactions occurring in the strip 3 of the MUSIC detector.
Left panel: sums of energy losses ($\sum_{1}^{16}\Delta E^{normalized}_i$, $\sum_{4}^{10}\Delta E^{normalized} _i$) are shown for events selected on an $^{87}$Rb incoming beam and a sharp increase in energy loss measured in the strip 3. The energy losses per strip were normalized to the averaged value of 12~MeV found for the $^{87}$Rb beam. The isolated region at $\sum_{4}^{10}\Delta E^{normalized}_i\sim82$ corresponds to ($\alpha,xn$) events, and scattering events are noticed below. Right panel: individual traces of calibrated energy losses $\Delta E$ along the strip number are shown for the unreacted beam (black curves), the ($\alpha,xn$) (solid red curves) and few ($\alpha,\alpha'$) at low angles (dotted red curves) reactions. The ($\alpha,xn$) events were obtained with a cut applied on the associated region in the {\itshape{PID}} plot (left). }\label{fig:Events}
\end{figure*}
\par The $^{87}$Rb stable beam delivered by the Argonne Tandem Linac Accelerator System ATLAS had a charge state of 17$^+$, an energy of 4.61(1)~MeV/u, and an average intensity of 3.7$\times10^4$~pps. The beam energy was measured from the time-of-flight value for three resonator pairs located upstream of the MUSIC detector. The beam purity was assessed from the energy losses in the entrance of the detector.  A single contaminant was observed with an intensity of about $1/7$ that of the requested beam. It was $^{51}$V$^{10^+}$ considering the magnetic rigidity matching  $^{87}$Rb$^{17^+}$ and the measured energy losses. The MUSIC detector was filled with pure $^4$He gas at 555~Torr held by Ti foils of 1.30(5)~mg/cm$^2$ thickness at the entrance and exit sides for the beam. The beam energy loss after passing through the entrance foil was measured to be 46.9(9)~MeV. The anode of MUSIC is segmented into 18 strips of equal width (1.578~cm) along the beam axis, see Fig.~1 of~\cite{MUSICsetup}. In the data-acquisition system of digital nature (DAQ), each channel was self-triggering, and the beam rate was monitored to ensure the DAQ stability. Less than 10~\% pile-up events were observed as expected for such a gaseous detector operated at rates of tens of kHz. In the data analysis, a cut was applied on the energy losses in the first two strips to fully separate the $^{87}$Rb beam events from  $^{51}$V contaminant events. 
\par Energy losses of $^{87}$Rb in the Ti foils and $^{4}$He gas were measured with a depleted silicon surface-barrier detector that was mounted downstream of MUSIC. Note that to correct for the pulse-height defects related to heavy ions in Si detectors~\citep{WILKINS1971381}, the energy calibration of the Si detector was performed in-beam by measuring the $^{87}$Rb beam at several energies ($ 1.0 - 4.6$ MeV/u) in the Si detector. Regarding the energy loss measurement, the beam energy was measured after the MUSIC detector:~first, only with the Ti foils but no gas, second the detector was gradually filled with gas where the beam energy at a given pressure was recorded. The  measured energy losses were reproduced, i.e~within 2.0~\%, by Monte-Carlo simulations using the mean value of the stopping powers of the ATIMA table in \texttt{LISE++}~\citep{lise} and of the tables in \texttt{SRIM}~\citep{srim}.\\
\par Overall, the beam energy losses per strip were of $11 - 13$~MeV with a resolution of 8~\% (FWHM) due to the strip spatial extension. In the active region, several reaction channels were energetically allowed:~the Rutherford and inelastic scatterings referred to as ($\alpha,\alpha'$), the ($\alpha,1n$) channel with Q$_{value}=-3.75$~MeV, the ($\alpha,p$) with Q$_{value}=-3.51$~MeV, and the ($\alpha, \gamma$) channel with Q$_{value}=+4.18$~MeV. Additionally, the ($\alpha,2n$) channel with Q$_{value}=-10.61$~MeV was opened at the highest  energies ($E_{c.m.}\geq10.87$~MeV). The dominant mechanism is ($\alpha,\alpha'$) with calculated cross sections in excess of 1~barn (estimates from {\texttt{LISE++}}). Monte-Carlo simulations of the set-up indicated that ($\alpha,p$) and ($\alpha,\gamma$) events would overlap with ($\alpha,xn$) events. But statistical model cross sections from the {\texttt{Talys}} code revealed that, at present energies, ($\alpha,xn$) reactions should be favoured, by several orders of magnitude ($5\times10^3 - 10^5$), over ($\alpha,p$) and ($\alpha, \gamma$) reactions. Possible contamination from the latter was thereby negligible compared to the statistical uncertainty ($1 - 25$~\%) of the measured cross sections. Simulated energy losses of $^{89, 90}$Y recoils differ by less than a hundred keV:~this is far below the detector energy resolution. Hence, the $2n$ exit channel could not be isolated from the $1n$ exit channel. The measured inclusive $^{87}$Rb($\alpha,xn$) cross sections were compared with statistical model cross sections. This allowed us to extract their respective contributions at $E_{c.m.}\geq10.87$~MeV.
\par  The {\itshape{PID method}} was first employed to identify ($\alpha,xn$) and scattering events which occurred in each strip of the MUSIC detector. This is illustrated for the reactions taking place in the strip 3 on the left of Figure~\ref{fig:Events}. 
As detailed previously, two conditions were required:~energy losses measured in the first two strips corresponded to $^{87}$Rb, and energy losses measured in the strip 3 were higher than the ones in the strip 2. Two isolated regions are noticed but the one at higher energies is unambiguously associated to ($\alpha,xn$) events considering $Z$ of the respective recoils. Then, the {\itshape{Traces method}} was used to count ($\alpha,xn$) events. Such traces of ($\alpha,xn$) reactions occurring in the strip 3 are shown with the solid red curves on the right of Figure~\ref{fig:Events}. A third condition was used here with a cut applied on the ($\alpha,xn$) region (Figure~\ref{fig:Events} left). Several traces of unreacted beam events (black curves) and of beam-like  scattering events at low angles (dotted red curves) are also presented on the right of Fig.~\ref{fig:Events}. Similarly to~\cite{Ong22}, scattering reactions at large angles were observed with a higher variation of $\Delta E$ than ($\alpha,xn$) reactions.

\subsection{Results}
\par
\begin{figure}[ht!]
\includegraphics[scale=0.47]{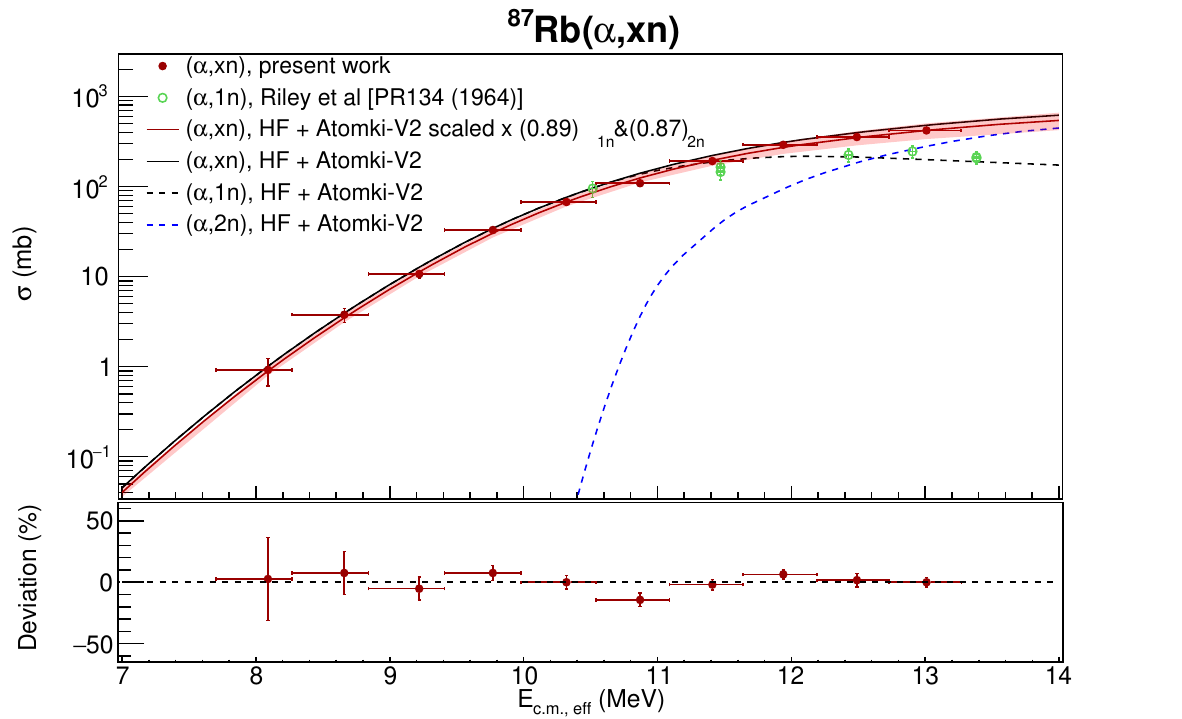}
\caption{Comparison of $^{87}$Rb($\alpha,xn$) cross sections as obtained from the present measurement and from the statistical model estimates. The common abscissa is the effective center-of-mass energy ($E_{c.m.,eff}$).
Upper panel:~experimental data are shown as red points. The ($\alpha,xn$) calculated cross sections including the $\alpha$OMP {\texttt{Atomki-V2}}~\cite{atomkiv2} (black curve) are dominated by the ($\alpha,1n$) channel (dotted black  curve) at low energy $E_{c.m., eff}<11$~MeV.  The ($\alpha,2n$) channel (dotted blue curve) starts to contribute ($>1$~\%) at $E_{c.m., eff}>10.8$~MeV, and dominates at high energy $E_{c.m., eff}>13.5$~MeV. The calculations scaled to experimental data are shown (dotted red curve). This corresponds to the best fit ($\chi^{2}/ndf=0.92$) of the weighted sum of the ($\alpha,1n$) and ($\alpha,2n$) calculations to measurements:~the deviations are of 0.89(9) and  0.87(14), respectively. Statistical model estimates, only 10~$\%$ higher than measurements, well reproduce the $^{87}$Rb($\alpha,xn$) cross sections at $E_{c.m., eff}=8 - 13$~MeV. The scaled calculations allow to assess cross sections at lower astrophysical energies (T$\leq$3.7~GK). Uncertainties contributions of the measured cross sections are detailed in the text and Table~\ref{tab:cs} (Appendix). The coloured band corresponds to the 3$\sigma$ uncertainty of the fit for the scaled calculations. Past work on $^{87}$Rb($\alpha,1n$)$^{90}$Y by~\cite{Riley1964} is presented (green points). Lower panel:~deviations between scaled calculations and measured data scatter around $0\pm10$~\%, and are not energy dependent.}\label{fig:cs}
\end{figure}

 The total cross sections of the $^{87}$Rb($\alpha,xn$) reactions are presented along the measured effective center-of-mass energies ($E_{c.m., eff}$) of $8.09 - 13.01$~MeV in Figure~\ref{fig:cs} and detailed in Table~\ref{tab:cs} (Appendix).
To determine $E_{c.m., eff}$, the center-of-mass energies deduced from the measurement of the beam energy losses in the MUSIC detector were corrected for the thick-target yield~\citep{PhysRevC.104.035804}. Shown with the red points, horizontal (energy) uncertainties of the measured cross sections correspond to the spatial extension of the strip and vertical uncertainties include both statistical and systematic contributions. Note that at low energies, the systematic contribution strongly increased because the ($\alpha,1n$) and ($\alpha,\alpha'$) channels became more entangled in both the {\itshape{PID method}} and {\itshape{Traces method}}. Due to the too small difference in $Z$ between Rb and Y relative to the total atomic number, the above effect puts a low-energy limit on the present set-up to assess ($\alpha,xn$) weak $r$-process cross sections.
\par The measured cross sections are compared to statistical model cross sections in Figure~\ref{fig:cs} (upper panel). Calculated ($\alpha,1n$) and ($\alpha,2n$) cross sections were obtained with the {\texttt{Talys}} code~\citep{talys1,talys2} using the $\alpha$OMP {\texttt{Atomki-V2}}~\citep{atomkiv2}. The results, shown as the dotted black and blue curves, respectively, were summed up to obtain the black curve representing the ($\alpha,xn$) cross sections (black curve). A fit was then performed between the measurement and the weighted sum of the ($\alpha,1n$) and ($\alpha,2n$) calculations, and the best result (dotted red curve) was obtained for a relative deviation of 0.89(9) and  0.87(14), respectively. The observed deviations were found consistent between the two exit channels, i.e.~they are equal within uncertainties. This was expected since both cross sections are governed by the same $\alpha$OMP. It should be noted that using a scaling factor per exit channel or a common factor of 0.88 resulted in negligible changes in the associated reaction rate. This new experimental result supports the use of $\alpha$OMP {\texttt{Atomki-V2}} to reliably predict ($\alpha,xn$) cross sections at astrophysical energies in weak $r$-process. The results of the $^{87}$Rb($\alpha,1n$)$^{90}$Y reaction measured at high energy by~\cite{Riley1964} are also shown in Figure~\ref{fig:cs} (green points). Agreeing with the present work at $E_{c.m., eff}=10.5$~MeV where the ($2n$) exit channel is suppressed, it confirms the robustness of statistical model estimates based on $\alpha$OMP {\texttt{Atomki-V2}}.
\par  Once scaled, the ($\alpha,xn$) calculated cross sections were found to differ only by $\pm 10$~\% in comparison to experimental data, see Fig.~\ref{fig:cs} (lower panel). No systematic trend was noticed along $E_{c.m., eff}$. The energy dependence of the experimental data was found to be properly reproduced by statistical model estimates. Therefore, information at energies lower than measured values could be assessed by using the derived constant scaling factor of 0.89(9) to the lowest energies for the $^{87}$Rb($\alpha,1n$)$^{90}$Y reaction of interest.
\par
\begin{figure}[ht!]
\begin{center}
\includegraphics[scale=0.47]{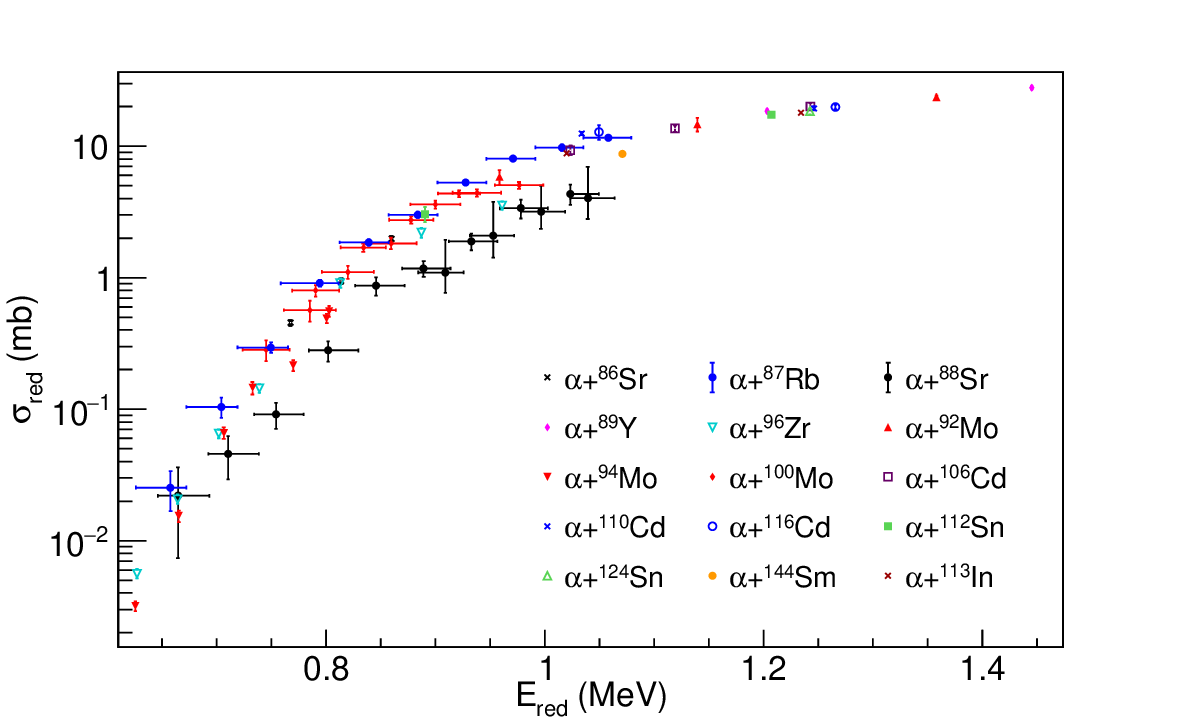}
\end{center}
\caption{Evolution of the reduced cross section ($\sigma_{red}$) as a function of the reduced energy ($E_{red}$) for $A\approx86-144$ nuclei produced in $\alpha$-induced reactions at low energy from~\citep{redSigma, ref86Sr, Kiss2021, PhysRevC.78.025804, Ong22, PhysRevC.109.065805}. Present results on $^{87}$Rb (blue circles) follow the apparent trend. \label{fig:cs_syst}}\end{figure}

 Looking at the reduced cross section ($\sigma_{red}$) and energy ($E_{red}$)~\citep{Gomes_PRC2005_reduced, redSigma} allows to directly compare total cross sections of charged-particle reactions whatever the beam, target and energy are. These two parameters are given by $\sigma_{red}=\frac{\sigma}{(A_{beam}^{1/3}+A_{target}^{1/3})^2}$ and $E_{red}=\frac{E_{c.m.}(A_{beam}^{1/3}+A_{target}^{1/3})}{Z_{beam}Z_{target}}$. A common trend was observed for $\alpha$-induced reactions around $A\sim100$ in~\cite{ redSigma}. This is illustrated in Figure~\ref{fig:cs_syst} where the result of the $\alpha+^{87}$Rb reaction fits well to this trend, similarly to several nuclei with $N=50$. If confirmed to be the case for even more nuclei, this trend may also be of use to infer weak $r$-process cross sections not yet measured.

\section{Astrophysical implications} \label{sec:ast}
\subsection{Thermonuclear reaction rates}
\par The $^{87}$Rb($\alpha,1n$)$^{90}$Y cross sections were measured at the energies corresponding to the Gamow temperatures of 3.7--7.6~GK. Calculations of the associated thermonuclear reaction rate were performed with the code {\tt{EXP2RATE}}~\cite{refExp2Rate}. They included the cross sections from statistical model estimates based on the Atomki-V2 $\alpha$OMP~\cite{atomkiv2} scaled by the constant factor of 0.89(9) deduced from the experiment. Recommended values of the $^{87}$Rb($\alpha,1n$)$^{90}$Y reaction rate at temperatures where the weak $r$-process impacts the nucleosynthesis in $\nu$-driven ejecta of CCSNe are reported in Table~\ref{tab:rate}.  
\begin{table}[ht!]
\caption{Low, recommended, and high thermonuclear rates of the $^{87}$Rb($\alpha,1n$)$^{90}$Y reaction, in units of cm$^3$ mol$^{-1}$ s$^{-1}$, for temperatures relevant for the weak $r$-process .\label{tab:rate} }
\begin{ruledtabular}
\renewcommand{\arraystretch}{1.1}
\begin{tabular}{c|ccc}

$T$ (GK)& Low & Recommended & High \\
\hline
1.6  &6.58$\times$10$^{-14}$&1.13$\times$10$^{-13}$& 1.93$\times$10$^{-13}$\\
2.0 & 6.66$\times$10$^{-11}$&1.09$\times$10$^{-10}$ & 1.80$\times$10$^{-10}$ \\
2.5 &3.51$\times$10$^{-8}$& 5.27$\times$10$^{-8}$  &7.91$\times$10$^{-8}$\\
3.0  &3.95$\times$10$^{-6}$&5.41$\times$10$^{-6}$ &7.42$\times$10$^{-6}$\\
3.5 & 1.61$\times$10$^{-4}$&2.05$\times$10$^{-4}$ & 2.62$\times$10$^{-4}$\\
4.0& 3.20$\times$10$^{-3}$ &3.88$\times$10$^{-3}$ &4.70$\times$10$^{-3}$\\
4.5  &3.75 $\times$10$^{-2}$&4.40$\times$10$^{-2}$& 5.15$\times$10$^{-2}$\\
5.0  &2.93$\times$10$^{-1}$&3.37$\times$10$^{-1}$ & 3.87$\times$10$^{-1}$ \\
5.5 &1.95 &2.17  &2.39\\
6.0  & 7.45  &8.39  &9.44 \\
\end{tabular}
\end{ruledtabular}
\end{table} The lower and higher limits of the reaction rate were derived from  uncertainties of the scaling factor that range from $\sim10\%$ at measured energies ($E_{c.m.}\geq8$~MeV) to a factor 2 at the threshold energy of the reaction ($E_{c.m.}=3.8$~MeV). The $^{87}$Rb($\alpha,1n$)$^{90}$Y reaction rate was found to be exceeded by the $^{87}$Rb($\alpha,2n$)$^{89}$Y reaction rate at $T>8$~GK. In similar calculations to assess the rate of the latter reaction, the statistical model cross sections were weighted by the constant scaling factor found for the $2n$ exit channel (0.87(14)). Note that the tabulated rates of the two reactions for each value of the default temperature grid of the \texttt{Talys} code are given in Table~\ref{tab:rate2} (Appendix).
\par
\begin{figure}[ht!]
\begin{center} 
\includegraphics[scale=0.45]{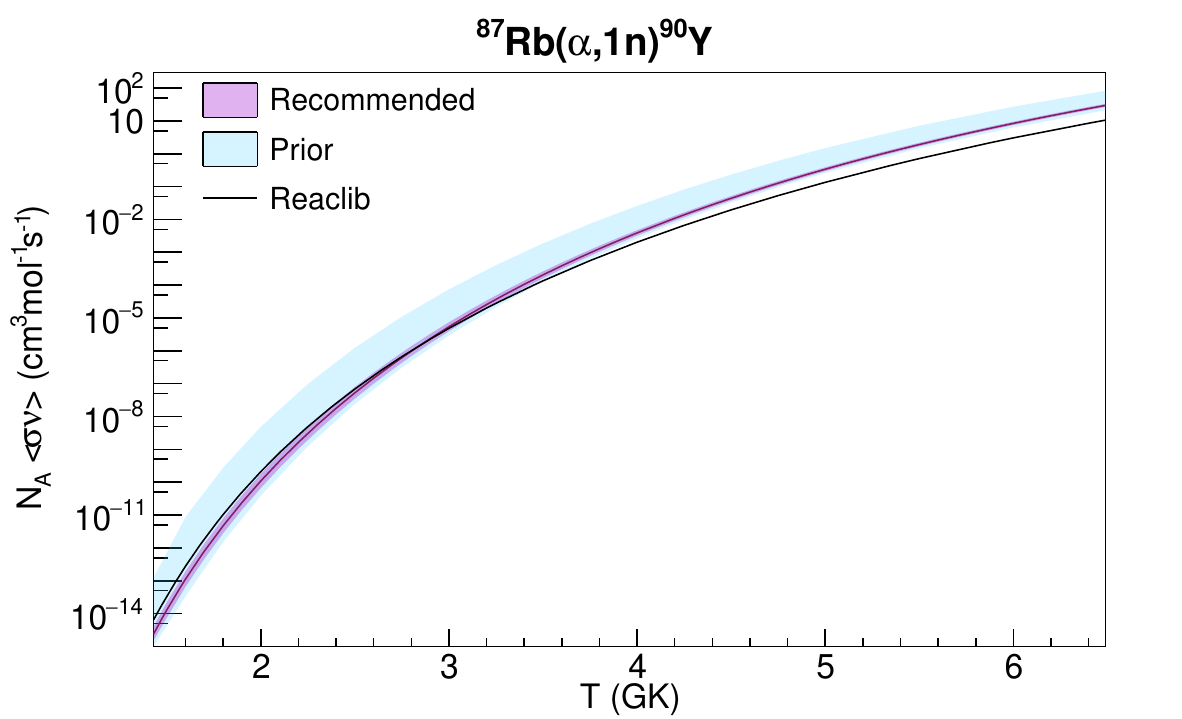}
\end{center}
\caption{Evolution of the thermonuclear reaction rate of $^{87}$Rb($\alpha,1n$)$^{90}$Y as a function of temperature relevant for the weak $r$-process. The recommended reaction rate (violet curve) was determined using the code {\tt{EXP2RATE}} based on cross sections from \texttt{Talys+Atomki-V2} scaled by the constant factor of 0.89. Resulting uncertainties are $\sim65 - 15$~\% at $T\sim2-5$~GK. The prior rate, shown as the cyan band, was evaluated from \texttt{Talys} calculations based on standard $\alpha$OMPs (Figure~\ref{fig:RateFluc}, uncertainties of 100 - 10). The reaction rate from \texttt{ReaclibV2.2} (black curve) agrees with the rate here reevaluated (within a factor of $2$). \label{fig:rate}}\end{figure}

 The evolution of the recommended rate of the  $^{87}$Rb($\alpha,1n$)$^{90}$Y reaction as a function of temperature is shown in Figure~\ref{fig:rate} (violet curve).  
The status prior to the measurement, shown as the cyan band, was obtained from calculations which included standard $\alpha$OMPs (see Figure~\ref{fig:RateFluc}), low (high) limits corresponding to~\cite{aOMPdemetriou} (~\cite{aOMPnolte}). The precision of the $^{87}$Rb($\alpha,1n$)$^{90}$Y reaction rate is drastically increased  at $T\sim2-5$~GK, i.e.~current uncertainties based on experimental data are $\sim 65-15\%$ whereas prior uncertainties based on statistical model calculations are a factor of 100 - 10. The referenced rate from {\texttt{ReaclibV2.2}}~\citep{Cyburt2010} is also given in Figure~\ref{fig:rate} (black curve). At weak $r$-process temperatures, the two rates agree within a factor of 2. This is in line with the initial development  where the {\texttt{ReaclibV2.2}} rate was observed to be close to statistical model estimates
based on the {\texttt{Atomki-V2}} $\alpha$OMP (see Fig.~\ref{fig:RateFluc}).
\begin{figure*}[ht!]
\begin{center}
\includegraphics[width=0.48\textwidth]{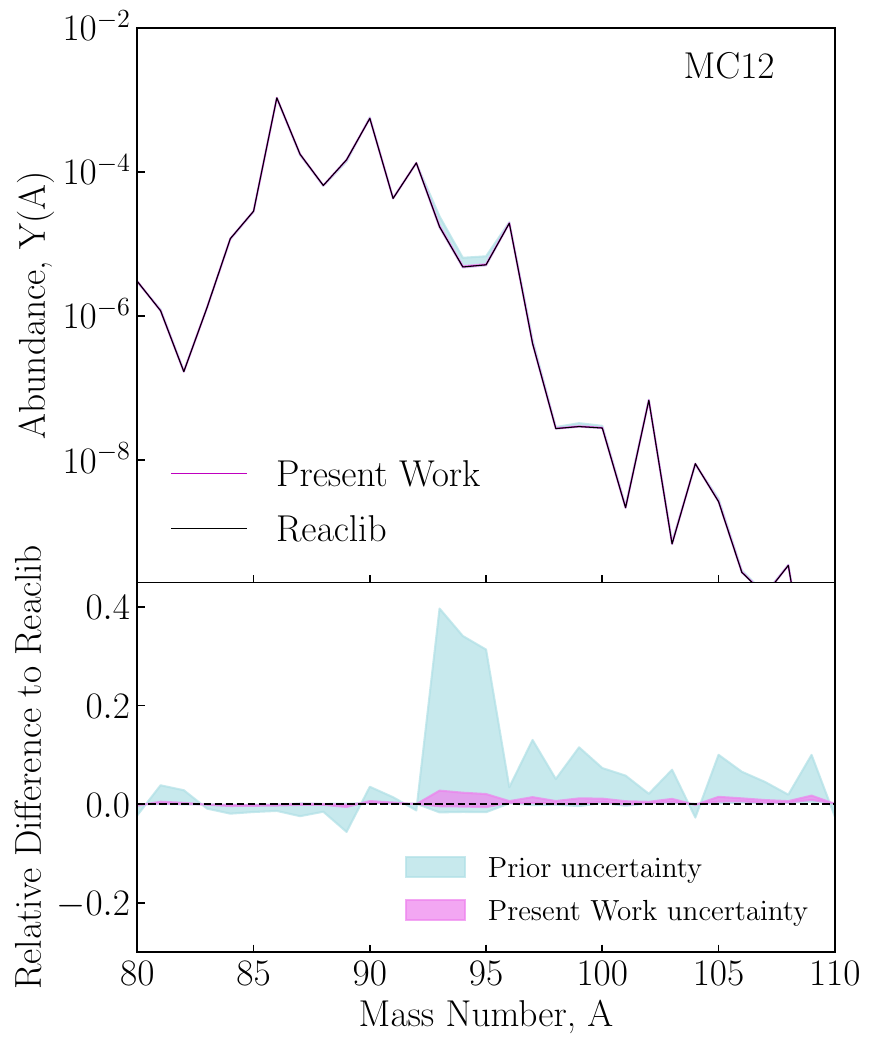}\includegraphics[width=0.48\textwidth]{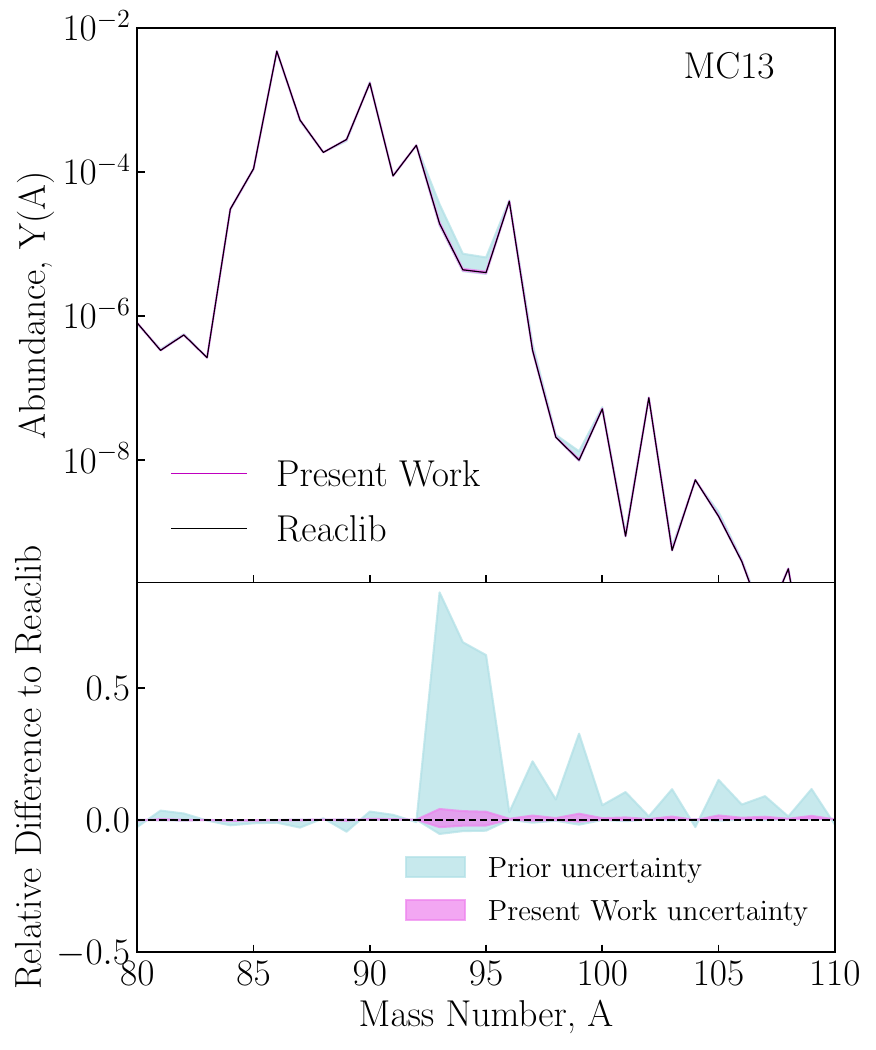}
\includegraphics[width=0.47\textwidth]{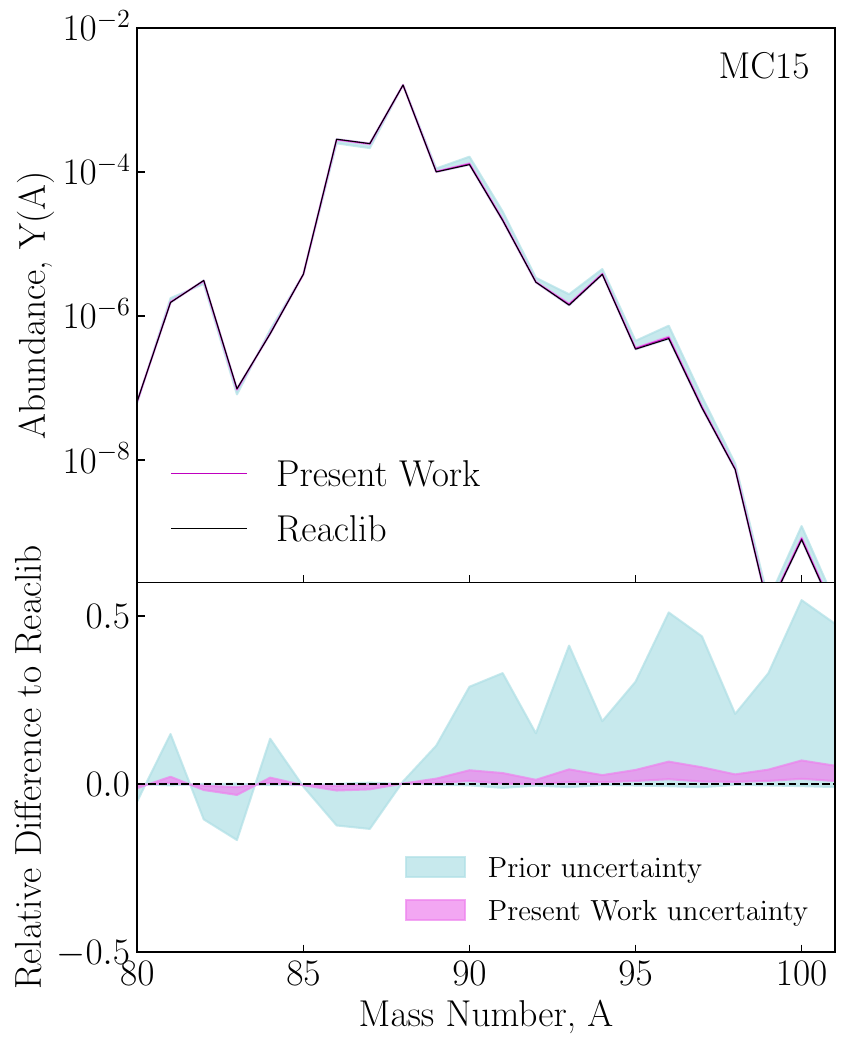}\includegraphics[width=0.48\textwidth]{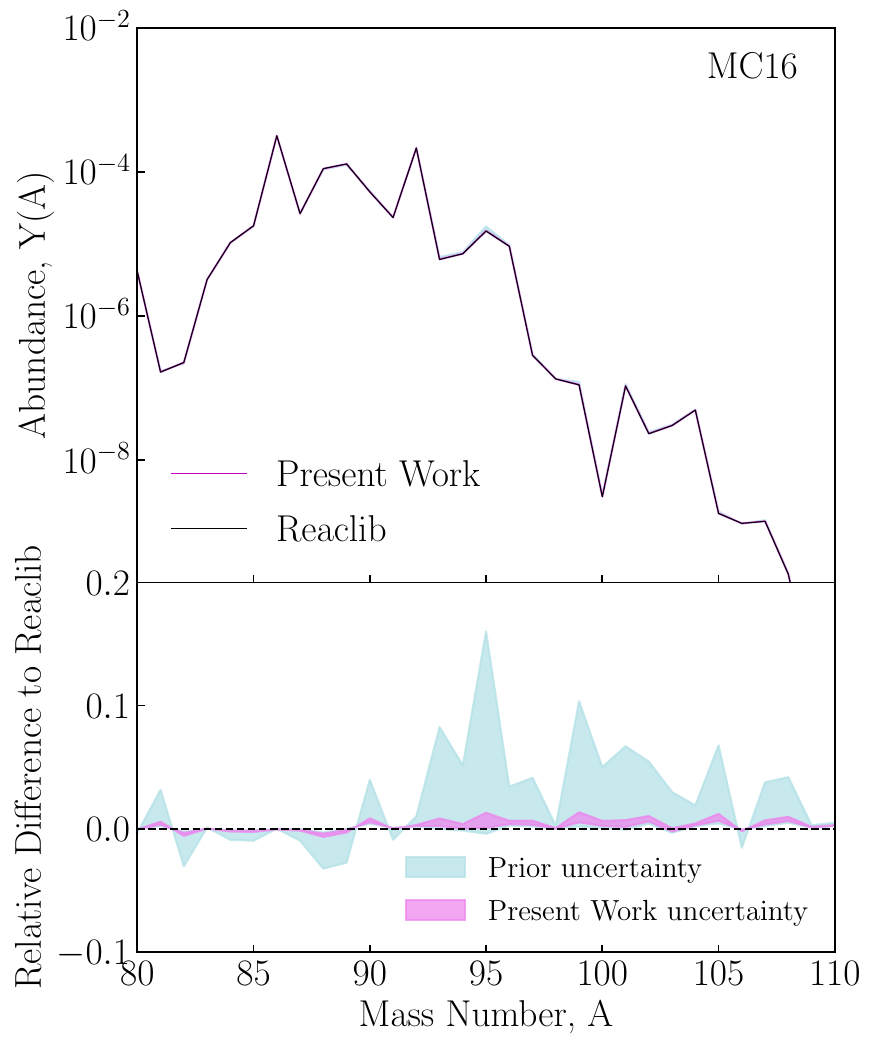}
\caption{Isotopic abundance patterns for four different sets of $\nu$-driven ejecta corresponding to the conditions MC12, MC13, MC15, and MC16 detailed in Table~\ref{MCccsneTab} (Appendix). The calculations use the setup of~\cite{Psaltis2022}. 
Each panel shows the abundance pattern as a function of mass number A for three cases: calculations using the {\texttt{ReaclibV2.2}}~$^{87}$Rb($\alpha,1n$)$^{90}$Y rate (black), the rate of the present work (magenta), and the previously recommended rate (cyan). 
The bottom panels illustrate the relative differences between each calculation and the one using the {\texttt{ReaclibV2.2}} rate. 
The new constrained reaction rate reduces the uncertainty in isotopic production for $A>90$. 
See the text for further details.
\label{fig:abundGlobal}}
\end{center}
\end{figure*} 
\subsection{Elemental abundances}
\par  The impact of the newly-constrained $^{87}$Rb($\alpha,1n$)$^{90}$Y reaction rate on the weak $r$-process abundances around $Z\sim40$ was studied via extensive nucleosynthesis calculations. We used the four thermodynamical conditions MC12, MC13, MC15 and MC16 from~\cite{Bliss2020} (see Table~\ref{MCccsneTab}), which have shown sensitivity to the $^{87}$Rb($\alpha,1n$)$^{90}$Y reaction rate in the works of~\cite{Bliss2020, Psaltis2022}, with the same reaction network setup as in~\cite{Psaltis2022}. With the exception of MC13, the electron fraction is high and two trajectories (MC15, MC16) have a high entropy ($s>100$~$k_B$ nucleon$^{-1}$).
\par Figure~\ref{fig:abundGlobal} shows the nucleosynthesis results for each of the aforementioned weak $r$-process thermodynamical conditions. Each panel displays the final isotopic pattern after 1~Gy using the {\texttt{ReaclibV2.2}} rate for the $^{87}$Rb($\alpha,1n$)$^{90}$Y reaction as a baseline case (black line). Additionally, we calculated the range of abundances using the prior $^{87}$Rb($\alpha,1n$)$^{90}$Y reaction rate based on standard $\alpha$OMPs (cyan shaded region). Furthermore, we repeated the calculations using the new experimental reaction rate (Table~\ref{tab:rate2}) together with its constrained uncertainty (magenta shaded region). The use of the experimental reaction rate for $^{87}$Rb($\alpha,1n$)$^{90}$Y reduces the uncertainty in the yields of the $A>90$ species to a few percent. By comparison, the use of the prior rate results in uncertainties of $\sim 50\%$ for some conditions.
\par
\begin{figure}[ht!]
\begin{center}
\includegraphics[width=0.48\textwidth]{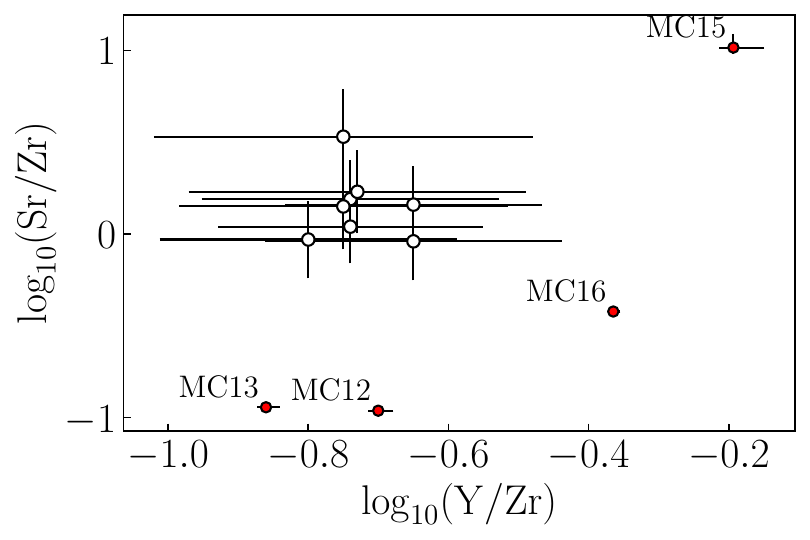}
\end{center}
\caption{Comparison between nucleosynthesis models and elemental abundance ratios of metal-poor stars from the compilation of~\cite{Psaltis2022}. 
The four astrophysical conditions are presented in Table~\ref{MCccsneTab} (Appendix). The error bars correspond to the uncertainty for the ratios based on the prior $^{87}$Rb($\alpha,1n$)$^{90}$Y reaction rate. 
The calculations with the new experimentally constrained rate are presented as the red points. The new uncertainty based on the present $^{87}$Rb($\alpha,1n$)$^{90}$Y reaction rate is smaller than the size of the points. The selected ejecta conditions fail to reproduce abundance observations in several metal-poor stars.\label{fig:abundLocal}}
\end{figure}

 We also compared the new nucleosynthesis results with the elemental abundance ratios -- Sr/Zr and Y/Zr --  of metal-poor stars that exhibit an excess of Sr - Ag relative to Solar values, as reported by \cite{Psaltis2022}. Despite showing some sensitivity to the uncertainty in the $^{87}$Rb($\alpha, 1n$)$^{90}$Y reaction rate, the four astrophysical conditions we investigated were unable to reproduce the observed abundance ratios, consistent with the findings of \cite{Psaltis2022}. However, as illustrated in Figure~\ref{fig:abundLocal},  our current experiment has significantly reduced the uncertainty in these abundance ratios. This improvement in one of the major nuclear physics uncertainties allows us to confidently rule out these conditions of the $\nu$-driven winds as potential candidates for reproducing the observed abundance ratios in these metal-poor stars.


\section{Conclusion}
\par The $^{87}$Rb($\alpha,1n$)$^{90}$Y reaction has been reported to impact $Z\sim40$ abundances which should be produced by the weak $r$-process ongoing in $\nu$-driven winds after core-collapse supernovae. The present work provides the first experimental insight into this reaction at astrophysical temperatures. The total cross sections of $^{87}$Rb($\alpha,xn$) reactions were measured at $T=3.7-7.6$~GK with the active gaseous target MUSIC. Measured values were found to be highly consistent with statistical model estimates based on the $\alpha$OMP {\texttt{Atomki-V2}}. This led to a robust re-evaluation of the thermonuclear rates of the $^{87}$Rb($\alpha,1n$)$^{90}$Y and $^{87}$Rb($\alpha,2n$)$^{89}$Y reactions. The experimental inputs indeed constrained, by two-to-three orders of magnitude, the $^{87}$Rb($\alpha,1n$)$^{90}$Y reaction rate at weak $r$-process temperatures, i.e.~the present uncertainties of $15 - 65$~\% are to be compared with the prior uncertainties of $10-100$. This test of the input of the statistical model calculations for cases at $N=50$ further supports the potential {\texttt{Atomki-V2}} in its predictive power of $\alpha$-induced reaction cross sections. In the near future, more and more constraints on ($\alpha,xn$) weak $r$-process reactions are expected from the ongoing experimental program with the MUSIC detector harvesting the neutron-rich beams available at the ATLAS and FRIB accelerators. 
\par The impact of the re-evaluated rate of the $^{87}$Rb($\alpha,1n$)$^{90}$Y reaction was assessed in detailed nucleosynthesis calculations for different $\nu$-driven trajectories of various thermodynamics conditions, introduced in previous impact studies~\citep{Bliss2020, Psaltis2022}. Under the investigated conditions, the experimental reaction rate reduces the production uncertainty of $A>90$ species to $\approx 10\%$. As expected, from previous studies, our comparison with observed abundance ratios did not yield any matches. However, it has helped eliminate these $\nu$-driven conditions as the candidates, narrowing down the possibilities for reproducing the peculiar abundance patterns in the Sr-Ag region observed in metal-poor stars in the Galaxy.

\section{Acknowledgments}
The authors thank the support of the ATLAS beam and detection physicists. This material is based upon work supported by the U.S. Department of Energy, Office of Science, Office of Nuclear Physics, under Contract No. DE-AC02-06CH11357. This research used resources of Argonne National Laboratory’s ATLAS facility, which is a DOE Office of Science User Facility. A.P. acknowledges support from the U.S. Department of Energy, Office of Science, Office of Nuclear Physics, under Award Number DE-SC0017799 and Contract Nos. DE-FG02-97ER41033 and DE-FG02-97ER41042. 
P.M. would like to acknowledge the support of NKFIH (grant K134197). S.B. acknowledges support from the Institute for Basic Science, South Korea (IBS), under the Grants No. IBS-R031-D1.
\bibliography{sample631}{}
\bibliographystyle{aasjournal}
\appendix
\section{Cross-section data}
\begin{table}[ht!]
\caption{Reported are the cross sections ($\sigma_{(\alpha,xn)}$) of the $^{87}$Rb($\alpha,xn$) reactions measured for a given effective center-of-mass energy  ($E_{c.m., eff}$). The uncertainty of $E_{c.m., eff}$ includes the contributions of measured energy losses and incident beam energy. The energy range mentioned covers the strip. Statistical and systematic uncertainties of $\sigma_{(\alpha,xn)}$ are given. }\label{tab:cs}
\begin{ruledtabular}
\renewcommand{\arraystretch}{1.3}
\begin{tabular}{lcc|ccc}
$E_{c.m., eff}$\tablenotemark{a} &
\textrm{Range}\tablenotemark{b} &Gamow temperature &\textrm{$\sigma_{(\alpha,xn)}$ (mb)}
&\multicolumn{2}{c}{\textrm{Uncertainties (\%)}}\\
\multicolumn{2}{c}{(MeV)} & (GK) & & statistical & systematic \\
\hline
13.01$^{+0.26}_{-0.28}$&[13.27, 12.73]& 7.6 &420(17) &1.01& 3.95\\
12.49$^{+0.24}_{-0.30}$&[12.73, 12.19]& 7.2 &355(19) &1.13& 5.37\\
11.94$^{+0.25}_{-0.30}$&[12.19, 11.64]& 6.7  &291(11) &1.28& 3.71\\
11.41$^{+0.23}_{-0.32}$&[11.64, 11.09]& 6.2 &192(8) &1.62& 4.32\\
10.87$^{+0.22}_{-0.33}$&[11.09, 10.54]& 5.8 &109(6) &2.19& 5.40\\
\hline
10.32$^{+0.22}_{-0.33}$&[10.54, 9.98]& 5.4&67.4(38) &2.81& 4.87\\
9.77$^{+0.21}_{-0.44}$&[9.98, 9.41]& 5.0 &32.9(20) &4.08& 4.62\\
9.22$^{+0.19}_{-0.38}$&[9.41, 8.84]& 4.5 &10.7(10) &7.24& 6.18\\
8.66$^{+0.18}_{-0.39}$&[8.84, 8.27]& 4.1&3.77(66) &12.3& 12.4\\
8.09$^{+0.18}_{-0.39}$&[8.27, 7.70]& 3.7 & 0.92(31) &25.0& 23.6\\
\end{tabular}
\tablenotetext{a}{At effective strip thickness corrected from the thick-target yield.}
\tablenotetext{b}{From entrance to strip exit.}
\end{ruledtabular}
\end{table}

 \section{Tabulated $^{87}$R\MakeLowercase{b}($\alpha$,\MakeLowercase{xn}) reaction rates}
\begin{table}[H]
\caption{Recommended, low, and high thermonuclear rates of the $^{87}$Rb($\alpha,1n$)$^{90}$Y  and $^{87}$Rb($\alpha,2n$)$^{89}$Y reactions, in units of cm$^3$ mol$^{-1}$ s$^{-1}$, along the default temperature grid of the \texttt{Talys} code.\label{tab:rate2} }
\begin{ruledtabular}
\renewcommand{\arraystretch}{1.1}
\begin{tabular}{c|ccc|ccc}
$T$ (GK)& Recommended &Low   & High &   Recommended & Low & High \\
&\multicolumn{3}{c|}{($\alpha,1n$)}&\multicolumn{3}{c}{($\alpha,2n$)} \\
\hline
0.10&$<10^{-30}$ &$<10^{-30}$ &$<10^{-30}$&$<10^{-30}$ &$<10^{-30}$ &$<10^{-30}$\\
0.15 &$<10^{-30}$ &$<10^{-30}$ &$<10^{-30}$&$<10^{-30}$ &$<10^{-30}$ &$<10^{-30}$\\
0.20 &$<10^{-30}$ &$<10^{-30}$ &$<10^{-30}$&$<10^{-30}$ &$<10^{-30}$ &$<10^{-30}$\\
0.25&$<10^{-30}$ &$<10^{-30}$ &$<10^{-30}$&$<10^{-30}$ &$<10^{-30}$ &$<10^{-30}$\\
0.3&$<10^{-30}$ &$<10^{-30}$ &$<10^{-30}$&$<10^{-30}$ &$<10^{-30}$ &$<10^{-30}$\\
0.35&$<10^{-30}$ &$<10^{-30}$ &$<10^{-30}$&$<10^{-30}$ &$<10^{-30}$ &$<10^{-30}$\\
0.40 &$<10^{-30}$ &$<10^{-30}$ &$<10^{-30}$&$<10^{-30}$ &$<10^{-30}$ &$<10^{-30}$\\
0.45&$<10^{-30}$ &$<10^{-30}$ &$<10^{-30}$&$<10^{-30}$ &$<10^{-30}$ &$<10^{-30}$\\
0.50 &$<10^{-30}$ &$<10^{-30}$ &$<10^{-30}$&$<10^{-30}$ &$<10^{-30}$ &$<10^{-30}$\\
0.60&$<10^{-30}$ &$<10^{-30}$ &$<10^{-30}$&$<10^{-30}$ &$<10^{-30}$ &$<10^{-30}$\\
0.70 &8.62$\times$10$^{-30}$ & 5.17$\times$10$^{-30}$ &1.47$\times$10$^{-29}$ &$<10^{-30}$ &$<10^{-30}$ &$<10^{-30}$\\
0.80& 2.36$\times$10$^{-26}$&1.41$\times$10$^{-26}$ &4.01$\times$10$^{-26}$ & $<10^{-30}$ &$<10^{-30}$ &$<10^{-30}$\\
0.90 &1.17$\times$10$^{-23}$ &7.02$\times$10$^{-24}$& 1.99$\times$10$^{-23}$ & $<10^{-30}$ &$<10^{-30}$ &$<10^{-30}$\\
1.0 &1.77$\times$10$^{-21}$ &1.06$\times$10$^{-21}$ & 3.01$\times$10$^{-21}$  & $<10^{-30}$ &$<10^{-30}$ &$<10^{-30}$\\
1.2& 2.99$\times$10$^{-18}$ &1.73$\times$10$^{-18}$ &5.16$\times$10$^{-18}$& $<10^{-30}$ &$<10^{-30}$ &$<10^{-30}$\\
1.4 &1.13$\times$10$^{-15}$ &6.55$\times$10$^{-16}$ &1.95$\times$10$^{-15}$& $<10^{-30}$ &$<10^{-30}$ &$<10^{-30}$\\
1.6 &1.13$\times$10$^{-13}$ &6.58$\times$10$^{-14}$& 1.93$\times$10$^{-13}$& 1.21$\times$10$^{-26}$ & 7.05$\times$10$^{-27}$& 2.06$\times$10$^{-26}$ \\
1.8 &4.73$\times$10$^{-12}$& 2.80$\times$10$^{-12}$ &7.97$\times$10$^{-12}$&  1.80$\times$10$^{-22}$ &1.07$\times$10$^{-22}$& 3.03$\times$10$^{-22}$ \\
2.0 &1.09$\times$10$^{-10}$ & 6.66$\times$10$^{-11}$& 1.80$\times$10$^{-10}$&  2.02$\times$10$^{-19}$ &1.24$\times$10$^{-19}$& 3.33$\times$10$^{-19}$ \\
2.25& 3.63$\times$10$^{-9}$ &2.31$\times$10$^{-9}$ &5.68$\times$10$^{-9}$&  2.31$\times$10$^{-16}$ &1.47$\times$10$^{-16}$& 3.61$\times$10$^{-16}$ \\
2.5 & 5.27$\times$10$^{-8}$ &3.51$\times$10$^{-8}$ &7.91$\times$10$^{-8}$&  6.60$\times$10$^{-14}$ &4.40$\times$10$^{-14}$& 9.90$\times$10$^{-14}$ \\
2.75 & 6.84$\times$10$^{-7}$ & 4.79$\times$10$^{-7}$& 9.76$\times$10$^{-7}$&  6.88$\times$10$^{-12}$ &4.82$\times$10$^{-12}$& 9.81$\times$10$^{-12}$ \\
3.0 &5.41$\times$10$^{-6}$ & 3.95$\times$10$^{-6}$ &7.42$\times$10$^{-6}$ & 3.37$\times$10$^{-10}$ &2.46$\times$10$^{-10}$ &4.62$\times$10$^{-10}$ \\
3.25& 3.92$\times$10$^{-5}$ &2.98$\times$10$^{-5}$ &5.15$\times$10$^{-5}$& 9.22$\times$10$^{-9}$ &6.78$\times$10$^{-9}$ &1.21$\times$10$^{-8}$ \\
3.5 &2.05$\times$10$^{-4}$ & 1.61$\times$10$^{-4}$& 2.62$\times$10$^{-4}$& 1.60$\times$10$^{-7}$ &1.25$\times$10$^{-7}$ &2.04$\times$10$^{-7}$ \\
3.75 &1.00$\times$10$^{-3}$ &8.07$\times$10$^{-4}$& 1.24$\times$10$^{-3}$& 1.92$\times$10$^{-6}$ &1.54$\times$10$^{-6}$ &2.39$\times$10$^{-6}$ \\
4.0 &3.88$\times$10$^{-3}$& 3.20$\times$10$^{-3}$ &4.70$\times$10$^{-3}$& 1.72$\times$10$^{-5}$ &1.41$\times$10$^{-5}$ & 2.09$\times$10$^{-5}$ \\
4.25 &1.42$\times$10$^{-2}$ &1.19$\times$10$^{-2}$ &1.69$\times$10$^{-2}$& 1.21$\times$10$^{-4}$& 9.99$\times$10$^{-5}$ & 1.45 $\times$10$^{-4}$\\
4.5 &4.40$\times$10$^{-2}$ &3.75$\times$10$^{-2}$& 5.15$\times$10$^{-2}$& 6.90  $\times$10$^{-4}$ &5.87$\times$10$^{-4}$ & 8.09 $\times$10$^{-4}$\\
4.75 &1.30$\times$10$^{-1}$& 1.12$\times$10$^{-1}$& 1.50$\times$10$^{-1}$&  3.32$\times$10$^{-3}$ & 2.85$\times$10$^{-3}$& 3.98$\times$10$^{-3}$ \\
5.0 &3.37$\times$10$^{-1}$ &2.93$\times$10$^{-1}$ & 3.87$\times$10$^{-1}$& 1.38$\times$10$^{-2}$ & 1.15$\times$10$^{-2}$ & 1.62$\times$10$^{-2}$ \\
5.5 &2.17$\times$10$^{0}$ &1.95$\times$10$^{0}$  &2.39$\times$10$^{0}$ & 1.66$\times$10$^{-1}$& 1.45$\times$10$^{-1}$ &1.90 $\times$10$^{-1}$\\
6.0 &8.39$\times$10$^{0}$ & 7.45$\times$10$^{0}$  &9.44$\times$10$^{0}$ &1.35$\times$10$^{0}$ &1.18$\times$10$^{0}$ &1.53$\times$10$^{0}$\\
6.5& 3.03$\times$10$^{1}$  &2.70$\times$10$^{1}$ &3.39$\times$10$^{1}$& 7.98$\times$10$^{0}$ &7.10$\times$10$^{0}$ &8.94$\times$10$^{0}$  \\
7.0 &9.28$\times$10$^{1}$ &8.31$\times$10$^{1}$ &1.04$\times$10$^{2}$& 3.63$\times$10$^{1}$& 3.24$\times$10$^{1}$ &4.08$\times$10$^{1}$ \\
7.5 &2.48$\times$10$^{2}$& 2.23$\times$10$^{2}$ &2.77$\times$10$^{2}$& 1.32$\times$10$^{2}$ &1.18$\times$10$^{2}$ &1.47$\times$10$^{2}$\\
8.0 &3.39$\times$10$^{2}$ &3.05$\times$10$^{2}$& 3.73$\times$10$^{2}$& 3.99$\times$10$^{2}$& 3.58$\times$10$^{2}$& 4.40$\times$10$^{2}$ \\
8.5& 5.65$\times$10$^{2}$ &5.08$\times$10$^{2}$ &6.22$\times$10$^{2}$& 1.02$\times$10$^{3}$ &9.17$\times$10$^{2}$ &1.13$\times$10$^{3}$ \\
9.0 &8.46$\times$10$^{2}$ &7.60$\times$10$^{2}$& 9.31$\times$10$^{2}$& 2.29$\times$10$^{3}$ &2.05$\times$10$^{3}$& 2.52$\times$10$^{3}$ \\
9.5 &1.16$\times$10$^{3}$ &1.04$\times$10$^{3}$ &1.28$\times$10$^{3}$& 4.58$\times$10$^{3}$ &4.10$\times$10$^{3}$ &5.06$\times$10$^{3}$ \\
10 &1.48$\times$10$^{3}$& 1.33$\times$10$^{3}$ &1.63$\times$10$^{3}$& 8.33 $\times$10$^{3}$&7.48$\times$10$^{3}$ &9.17$\times$10$^{3}$ \\
\end{tabular}
\end{ruledtabular}
\end{table}
\section{Astrophysical conditions used in nuclear network calculations}

\begin{table}[H]
\caption{Astrophysical conditions of CCSNe $\nu$-driven winds and abundances at $Z\sim40$ impacted by the $^{87}$Rb($\alpha,1n$)$^{90}$Y reaction rate, as taken from~\cite{Bliss2020}:~the electron mass fraction $Y_e$, the entropy per baryon $s$, and the expansion timescale $\tau$. The uncertainty in the elemental ratios is based only on the variation of the $^{87}$Rb($\alpha,1n$)$^{90}$Y rate within its new experimental uncertainties.
}\label{MCccsneTab}
\begin{ruledtabular}
\renewcommand{\arraystretch}{1.2}
\begin{tabular}{c|ccc|ccc}
CCSNe   & \multirow{ 2}{*}{$Y_e$}
& $s$  & $\tau$   & \multirow{ 2}{*}{$\log_{10}$(Sr/Zr)}& \multirow{ 2}{*}{$\log_{10}$(Y/Zr)} & \multirow{ 2}{*}{$\log_{10}$(Nb/Zr)} \\
$\nu$-wind trajectory & &($k_B$ nucleon$^{-1}$) &(ms) & & &\\
\hline
12& 0.48& 85 & 9.7 & $-0.962 \pm 0.001$ & $-0.700 \pm 0.001$ & $-1.630^{+0.056}_{-0.085}$\\
13& 0.43&  64 & 35.9& $-0.944 \pm 0.001$ &  $-0.860 \pm 0.001$ & $-2.030^{+0.099}_{-0.182}$ \\
15&  0.48&  103 & 20.4 & $1.020^{+0.007}_{-0.006}$ & $-0.193 \pm 0.004$ & $-2.040^{+0.018}_{-0.023}$ \\
16&  0.49&  126 & 15.4 & $-0.422 \pm 0.001$& $-0.345 \pm 0.001$  & $-1.690^{+0.013}_{-0.016}$ \\
\end{tabular}
\end{ruledtabular}
\end{table}



\end{document}